\documentclass{aa}  

\usepackage{graphicx}
\usepackage{txfonts}
\usepackage{float}
\usepackage{ulem}  
\usepackage{graphicx}

\begin{document}

\title{Detection of binary companions below the diffraction limit with lucky imaging}
\titlerunning{Detection of binary companions below the diffraction limit}
\authorrunning{Cagigas et al.}

   \author{Miguel A. Cagigas,$^{1,2}$
   \and
R. Clavero\inst{1,2}
   \and
Manuel P. Cagigal\inst{3}
   \and
David Nespral\inst{1,2}
   \and
A. A. Djupvik\inst{4,5}
   \and
D. Jones\inst{1,2,4}
   \and
Pedro J. Valle\inst{3}
   \and
Vidal F. Canales\inst{3}
   \and
E. Soria\inst{1,2}
   \and
R. L\'opez\inst{1,2}
   \and
O. Zamora\inst{1,2}
   \and
\'A. Oscoz\inst{1,2}
   \and
J. Marco\inst{1,2}
          }

   \institute{Instituto de Astrof\'isica de Canarias, E-38205, La Laguna, Spain\\
   \email{miguel.cagigas@iac.es}
\and
Departamento de Astrof\'isica, Universidad de la Laguna, E-38206 La Laguna, Tenerife, Spain
   \and
Departamento de F\'isica Aplicada, Universidad de Cantabria, Avenida de los Castros 48, 39005 Santander, Spain
\and
Nordic Optical Telescope, Rambla Jos\'e Ana Fern\'andez P\'erez 7, 38711 Bre\~na Baja, Spain
\and
Department of Physics and Astronomy, Aarhus University, Ny Munkegade 120, 8000 Aarhus C, Denmark
             }

   \date{Received ; accepted }

 
\abstract{Binary stars are invaluable tools that can be used to precisely measure the fundamental properties of stars, to test stellar models, and further our understanding of stellar evolution. Stellar binarity may also play an important role in the formation and evolution of exoplanetary systems.}
{We provide a technique for resolving intermediate-separation binaries stars with medium-sized telescopes (i.e. diameter less than or equal to 2.5 metres) at wavelengths around 825 nm in the super-resolution range (i.e. below the limit defined by the Rayleigh criterion).}
{We combined two well-known algorithms that have been applied to reduce the halo in lucky imaging observations:\ COvariancE of Lucky Images (COELI) and the Lucky Imaging Speckle Suppression Algorithm (LISSA). We reviewed the fundamentals of both algorithms and describe a new technique called Lucky Imaging Super resolution Technique (LIST), which is optimized for peak highlighting within the first ring of the Airy pattern. To validate the technique, we carried out several observing campaigns of well-known binary stars  with the FastCam instrument (FC) on the 1.52 m Carlos S\'anchez Telescope (TCS) and 2.56 m Nordic Optical Telescope (NOT), both located at the
Observatorios de Canarias (OCAN).}
{The projected angular separation between objects was resolved by applying LIST to FC data taken with TCS and NOT, with a result below 0.15\arcsec{}. It can go down to approximately 0.05\arcsec{}, given the limitations of the detector plate scale. This is, to our knowledge, the first time that binary companions with such small angular separations have been detected using only lucky imaging at optical wavelengths.
The average accuracy achieved for the angular separation measurement is $16 \pm 2$ mas with  NOT and is $20 \pm 1$ mas with  TCS. The average accuracy obtained for the position angle measurement is $9.5 ^o \pm 0.3 ^o$ for NOT and $11 ^o \pm 2 ^o$ for  TCS.
We also made an attempt  to measure the relative brightnesses of the binary components, obtaining results that are compatible with literature measurements. Using this comparison, the $\Delta$m uncertainty obtained was 0.1 mag for NOT and 0.48 mag for TCS, although it should be noted that the measurements have been taken using slightly different filters.}
{Lucky imaging, in combination with speckle suppression and a covariance analysis, can allow the resolution of multiple point sources below the diffraction limit of 2-m class telescopes. However, it should be noted that measurements in the super-resolution regime are less sensitive than those above the first Airy ring.} 

   \keywords{Atmospheric effects -- Techniques: image processing -- Methods: statistical -- Methods: observational -- Binaries: visual -- Binaries: close
              }

   \maketitle
%

\section{Introduction}
\label{sec:Introduction}

The impact of atmospheric turbulence on the image quality of ground-based telescopes 
has been a recurring topic in astronomy for years. Typically, the quality of astronomical 
images obtained using large optical telescopes is constrained by the blurring effect caused 
by fluctuations in the refractive index within Earth’s atmosphere. Over time, various 
techniques have been developed to mitigate the effects of atmospheric turbulence on astronomical 
images, such as speckle interferometry \citep{{Weigelt1983},{2018speckle}}, speckle masking \citep{1983specmasking}, and adaptive optics \citep[AO;][]{{Hardy1998},{Tyson2017}}. These techniques have enabled us to approach the diffraction 
limit of large optical-infrared telescopes.

An alternative approach to addressing the aforementioned challenges is the lucky imaging (LI) technique, 
extensively discussed in \cite{Fried1978}. This technique involves capturing a series 
of short-exposure images and subsequently selecting the best ones based on their maximum peak 
intensity. Given the random nature of atmospheric fluctuations, it is anticipated that  the fluctuations will occasionally align fortuitously, resulting in a diffraction-limited  image.

Compared with other techniques, LI is particularly suitable for use with small- and medium-sized telescopes due to its simplicity and cost-effectiveness in terms of hardware requirements. Additionally, LI is able to  utilize reference stars that are fainter than those needed for the natural guide star AO technique.
One notable limitation of LI arises from the temporal evolution of atmospheric turbulence. 
For example, the de-correlation timescale (or atmospheric coherence time) associated with LI is approximately 30 ms at the Observatorios de Canarias (OCAN). Hence, the exposure times employed in LI must be shorter than this coherence time in order to `freeze' the atmospheric evolution. 
Under these conditions, the point spread function (PSF) becomes distorted, with its shape contingent on the ratio of the telescope diameter ($D$) to the Fried parameter ($r_0$), representing the atmospheric coherence length. The number of speckles manifested in the PSF  is roughly proportional to $(D/r_0)^2$ and randomly distributed within a circular region  in the image with an angular radius of  $1.22\lambda/r_0$.

This produces images with a peak surrounded by a number of speckles whose temporal average is commonly known as a halo. The peak is formed by the coherent part of the energy at the incoming wavefront added to an incoherent halo, while the surrounding speckle pattern is only due to the incoherent wavefront energy. It is important to note that the value of $r_0$ depends on the wavelength of observation, thereby impacting the quantity of speckles and the area they cover. Atmospheric conditions play a crucial role in real observations and (among various  other factors)  
are probably the most significant.

In a previous paper, we demonstrated that 
a proper image selection in LI can only be carried out when the height of the coherent peak is approximately twice that of the speckle mean intensity; that is, when $D/r_0 < 8$ \citep{Cagigas2013,Cagigal2016}. 
This condition is fulfilled for telescopes with diameters of up to 2.5 m, when observing in the $I$-band (between 800 to 900 nm). When $D/r_0$ exceeds this value, the presence of a brighter speckled halo 
compared to the coherent peak makes it impractical to select suitable images for analysis.

The LI technique has already had a significant impact in the detection of multiplicity in astronomical objects using medium-sized telescopes with angular resolution of around 0.15\arcsec{}--0.25\arcsec{} 
\citep[e.g.][]{2009A&A...499..729L,2012MNRAS.419..197R,2013MNRAS.429..859J,2017A&A...597A..47C,2022A&A...666A..16C}. In particular, it is relevant in the field of intermediate-separation binaries. 
For very close binaries, companions can frequently be detected via time-resolved photometry \citep{Prsa2022}, while wide binaries can
 often be identified through common proper motions \citep{Kervella2022}. 
Intermediate-separation binaries  have been generally more difficult to detect, requiring either lengthy 
campaigns of high-precision radial velocity monitoring \citep[e.g.][]{Jones2017} or  technically challenging  interferometric observations  to resolve the components \citep{Boffin2016}.  

In recent years, speckle interferometry has been employed with great success for the study of such intermediate-separation binaries \citep{Horch2016}, with instruments in active use at several medium-large class telescopes 
(e.g. WIYN 3.5 m telescope, \citealt{Howell2021}; 4.1-m SOAR telescope, \citealt{Ziegel2021}, \citealt{tokovinin24};  and the 8.1-m  Gemini telescopes, \citealp{Lester2021}, \citealp{scott21}).  
AO has also been used for such measurements in combination with speckle imaging \citep{2021AO-SOAR}, aperture-mask interferometry \citep{2016Kraus} or even the radial velocity technique \citep{2021Hirsch,2014Wang}. These observations can separate binaries down to the diffraction limit of the telescope employed, in some cases being complemented with long-term astrometric data such as Gaia \citep{2023Tokovininb,2022Salama} or even beating the detection limits of these space missions \citep{clark24}.

To detect potential companions in close proximity through high-resolution imaging observations with telescopes with a diameter of less than 2.5m using the LI technique, it is imperative to mitigate the adverse influence of the surrounding halo of incoherent energy. In this context, the COvariancE of Lucky Images (COELI) and Lucky Imaging Speckle Suppression Algorithm (LISSA) algorithms were developed. Both algorithms are based on the intensity speckle statistics, first described by \cite{1985Goodman} and applied by \cite{Canales99,Canales2001} to adaptive optics  and later confirmed experimentally by \cite{Steiger2022} and \cite{Bonse2023}. However, to develop these algorithms it is only necessary to take into account that fact that the intensity statistics in the lucky images is described by a Rician distribution.

\begin{figure}[H]
\begin{center}
        \includegraphics[width=0.55\columnwidth]{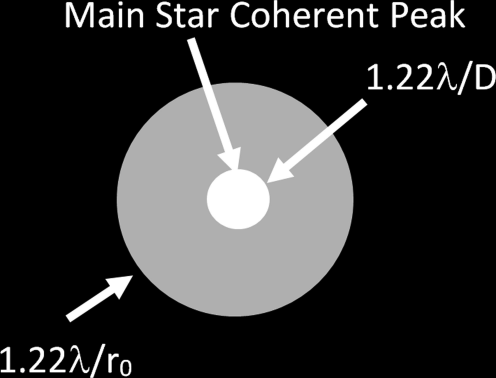}
    \caption{Area of applicability is found between the diffraction limit of the telescope $1.22 \lambda/D$ and the outer radius $1.22 \lambda/r_0$ for COLISSA and LF+COELI algorithms, and within the first Airy ring ($1.63 \lambda/D$) for the LIST algorithm.}
    \label{fig:fig1}
\end{center}
\end{figure}

The COELI technique is based on the estimation of the covariance between the intensity of the main star pixel and that of the remaining pixels throughout the LI cube, aimed at obtaining a two-dimensional (2D) covariance map. Wherever there is a companion, those pixels will show the highest covariance value \citep{Cagigal2017}.  Before applying the COELI algorithm, we may convolve the stack of lucky images with a Laplacian filter (LF) to improve the detectability of point objects. 

An enhancement of COELI consists of a reduction of the speckled halo in the LI cube that allows the achievement of better results. 
The halo can be removed by subtracting from the LI  cube the local standard deviation (STD) of each frame using the LISSA algorithm \citep{Cagigal2022}.
This allows us to compensate for the heteroscedasticity of the LI cube and, at the same time, to reduce the speckled halo around the host star, thus increasing the possibility of detecting potential companions. Both algorithms can be applied simultaneously (COLISSA) and optimized to detect companions in the area surrounding a host star. 
It is interesting to note that the LISSA algorithm can also be combined with any other technique. Therefore, LF+COELI and COLISSA are two powerful algorithms that have already been validated and can be applied either for extracting undetected faint companions from the background or to improve spatial resolution of images with detected companions to a main star. 
However, these techniques have thus far been  constrained by the diffraction limit of the telescope.
In particular, they have been applied in the region with inner radius equal to the limit defined by the Rayleigh criterion of $1.22 \lambda/D$   and with an outer radius of $1.22 \lambda/r_0$  ( the area covered by the halo; see Figure \ref{fig:fig1}).

In this paper, we present a combination of the aforementioned techniques with the innovation of applying them in the super-resolution regime, the so-called  Lucky Imaging Super resolution Technique (LIST). 
An image is said to have super-resolution when it is possible to detect two point objects separated by a distance less than the minimum distance given by the Rayleigh criterion ($1.22 \lambda/D$), which defines the diffraction limit of a telescope.
\cite{1985Toko} was among the first to realise the possibility of carrying out measurements below the diffraction limit in speckle observations. 
Then, \cite{Horch2006}  characterized binary stars in this regime with the Lowell-Tololo 61cm Telescope and  the WIYN 3.5m Telescope. Similar works, all of them using speckle observations, are presented by \cite{2011Horch} and \cite{2024Davidson} using the Differential Speckle Survey Instrument (DSSI).
In our case, we show that we can perform sub-diffraction-limited measurements, in what is known as the super-resolution regime, with telescopes where $D/r_0 < 8 $ is satisfied, using only optical LI measurements and post-processing analysis.
This extends the range of resolution for measuring separations and position angles in binary stars that can currently be achieved using LI alone.
By applying the LIST algorithm, results comparable to other techniques can be obtained with far more modest resources in terms of the instrumental and technological demands.

To demonstrate the applicability of the proposed technique,  the algorithm was applied to a set of experimental LI images taken with FastCam \citep{Oscoz2008} in the $I$-band by the 1.52 m Carlos S\'anchez Telescope (TCS) and 2.56 m Nordic Optical Telescope (NOT), both located at the OCAN. The  experimental set-ups and observations are detailed in Sects. \ref{sec:experiment} and  \ref{sec:observations}, respectively. The LIST algorithm is described in Sect. \ref{sec:Peak}.  This technique has been explored using synthetic binary-star cubes created from images of a single star obtained from real observations. The results for different separations and different relative intensities are shown in Sect. \ref{sec:simulations}. Finally, in Sect. \ref{sec:data}, the algorithm is applied to observations of
several binary objects with known orbits in the range under study.

\section{Experimental set-up}
\label{sec:experiment}

\begin{table*}[h]
 \caption{Binary stars observed  with separation at or below the resolution limit of  2.56 m NOT or 1.52 m TCS.
}

\label{tab:binaries}

\setlength{\tabcolsep}{15pt} 
\renewcommand{\arraystretch}{1.2} 
\begin{tabular}{ccccccc}
 
  \hline\hline
                WDS                         & Discover Comp  & HIP  & Date/Time  & Telescope & Magnitudes \\
        ($\alpha$,$\delta$,J2000.0) &                &      &       &         &  Pri  Sec     \\
        (1) &    (2)                &        (3)            &   (4)  &  (5)   & (6)          \\
\hline
08585+3548 & COU 1897   & 44064           &  2024-01-24T03:35    & NOT & 6.84,8.73\\
08585+3548 & COU 1897   & 44064           &  2024-03-17T22:39    &TCS & 6.84,8.73\\
10083+3136 & KUI 48AB   & 49658           &  2024-01-24T04:38    &NOT  & 6.90,7.20\\
10083+3136 & KUI 48AB   & 49658           &  2024-02-28T23:56    &TCS  & 6.90,7.20  \\
10116+1321 & HU 874     & 49929       &  2024-01-24T04:05    &NOT  & 6.90,7.87 \\ 
10116+1321 & HU 874     & 49929       &  2024-02-28T23:24    &TCS   & 6.90,7.87 \\
15245+3723 & CHR 181Aa,Ab & 75411     &  2023-05-10T01:38    &NOT  & 4.31,- \\
15245+3723 & CHR 181Aa,Ab & 75411     &  2023-04-12T03:38    &TCS  & 4.31,- \\
\hline
\end{tabular}
\tablefoot{The data for these objects have been used to test the LIST algorithm.
All objects were observed at zenith distances of less than 25$\degr{}$ and with $\lambda=825 nm $ and $\varDelta\lambda=120 nm$.The first three columns (1)-(3), are different designations for the objects, (4) is the date and time in UT of the observations with (5) the telescope used. In (6) the magnitudes from the WDS catalogue are listed. Note: for  object 15245+3723, only the magnitude of the main object is given because the source  this catalogue is based on does not include both measurements.
}
\end{table*}

FastCam (FC) is an instrument that was developed in 2006 by the Instituto de Astrof\'isica 
de Canarias (IAC) and Polytechnic University of Cartagena with the objective of 
obtaining very high spatial resolution images in the visible using ground-based telescopes and employing the LI technique \citep{Oscoz2008}. 
The IAC has since assumed its operation and it is now both a commonly-used instrument on  TCS and a visiting instrument on NOT. During its lifetime, the FC instrument has incorporated various detectors and has been continuously enhanced in terms of both hardware and software. The observations presented here were obtained with an Andor  EMCCD-type detector (iXon Ultra 888 model) with a 1024x1024 sensor with very fast readout speed (30 MHz).

The pixel size of this detector is 13 $\mu$m, which corresponds to a plate scale of 35 mas/pixel on  TCS and 25 mas/pixel on NOT. Obviously, if we want to use image processing techniques to obtain super-resolution, the image sampling has to be high enough to be able to distinguish such close point objects.
 LIST clears the area below the first Airy ring  ($1.63 \lambda/D$) by reducing the full width at half maximum (FWHM) of the Airy disk to less than a pixel.
Therefore the area inside the first ring has to be sampled with a minimum radius of 3-4 pixels to be able to resolve a secondary object in the super-resolution regime.
To meet this condition, it is essential that the plate scale is such that it produces this configuration which depends on the optical system used (telescope diameter and focal distance), the pixel size of the detector and the observed wavelength.

Observations were performed with a Johnson-Bessel $I$-band filter. The convolution of the filter transmission curve with the quantum efficiency of the detector results in a centroid wavelength of 825 nm and a bandwidth of 120 nm.
Using this effective central wavelength, the diffraction limit for the 1.52m TCS and 2.56m NOT is 0.136\arcsec{} (3.9 pixels) and 0.081\arcsec{} (3.2 pixels), respectively. For the calculation of the position of the first Airy ring, 0.183\arcsec{} (5.2 pixels) was obtained for  TCS and 0.108\arcsec{} (4.3 pixels) for  NOT. This configuration is optimal for detecting objects in the super-resolution regime with LIST, allowing us to extend the range for the detection of multiple astronomical point sources below separations of 0.15" and down to about 0.05", using only LI with FC and the aforementioned telescopes.

\section{Observations and data reduction}
\label{sec:observations}

This paper presents the results obtained by applying the LIST algorithm to a set of observations of astronomical objects taken during several observing campaigns in 2023 and 2024 using the FC instrument installed on  TCS, in addition to NOT.
Our strategy  was to take LI observations of known visual binaries stars. 
We selected objects for which the literature orbit of the system predicts a separation in the range near the diffraction limit for  NOT, but then in the super-resolution regime of  TCS due to its smaller diameter.
By having independent observations with two telescopes for the same objects, we are able to rule out systematic errors that falsify the detection of the components below the diffraction limit. Thus, we can check the possibility of detecting binaries in the super-resolution regime with the technique described.

We selected a number of visual binary stars from the Washington Double Star Catalog (WDS)  
with apparent angular separations on the order of (or less than) the diffraction limit at the time of observation. The next requirement in the selection was that these objects also be listed in the Sixth Catalogue of Orbits of  Visual Binary Stars  \citep[henceforth, Sixth Catalogue;][]{Hartkorpfroo1}. We selected binaries whose orbital elements have been
determined with a precision of grade 1 = definitive, 2 = good, or 3 = reliable, according to the evaluation criteria described in the last catalogue.  
Using these orbital elements and their errors, we calculated the expected ranges of the ephemerides on the observing nights to compare them with the measurements obtained from the observations after applying LIST (Sect. \ref{sec:data}).
With the current configuration on both telescopes (and because the plate scale is smaller than for past FC set-ups), we found that at least one star of magnitude $m_{V}$=11 (with good meteorological conditions and seeing between 0.8"--1.2") is required in order to obtain adequate statistics for the speckle distribution and to be able to make a good image selection and re-centering. However, to test the method presented here, we chose to select brighter stars of $m_{V}<7$, as it is expected that observations in the super-resolution regime are less sensitive both in terms of limiting magnitude and magnitude difference than those above the diffraction limit.  Future works will extend the method to fainter objects with companions to test whether a limiting magnitude comparable to observations above the diffraction limit can be reached.
In addition, data were obtained of the single star, HIP 52457  \citep[$m_{V}=5.07$;][]{Hog2000}, which were then used to perform simulations of multiple objects at different separations to be subsequently analysed with the LIST algorithm proposed here (Sect. \ref{sec:simulations}).

We carried out the observations with FC over five different nights. On these nights, conditions varied but were generally clear with seeing changing between 1$^{\prime\prime}$ and 2$^{\prime\prime}$. Using  TCS at the  Teide Observatory (OT, based on its initials in Spanish), WDS J15245+3723Aa,Ab, along with the single star HIP 52457, were observed on 2023 April 12. The same binary, WDS J15245+3723Aa,Ab, was also observed on the night of 2023 May 10 with  NOT located at Roque de los Muchachos Observatory (ORM, again based on its initials in Spanish). On the night of 2024 January 24, three more close double stars were observed with  NOT (WDS 10083+3136 AB, WDS 10116+1321, WDS 08585+3548). Of these objects, the first two were observed with  TCS on 2024 February 28, while the third was observed on 2024 March 17. All objects were observed at zenith distances of less than 25$\degr{}$.

In Table \ref{tab:binaries}, we provide information relating  to each of the observed binary stars with the following column information: (1) WDS number (this also gives the right ascension and declination of the object in J2000.0 coordinates); (2) discoverer designation and components involved; (3)  \textit{Hipparcos} catalogue (HIP) number \citep{hipparcos}; (4)  observation date and time in hours UT taking the midpoint of the observation; (5)  telescope of observations; (6) 
 V magnitude of the primary and secondary component as listed in the WDS catalogue. We give these magnitudes as a reference point but it should be noted that our observations are made in the $I$-band, where the difference in magnitudes may vary.

On each observing night, the plate scale and detector orientation were checked using visual binaries with separations greater than 2$^{\prime\prime}$ selected using the same criteria as we applied for the other objects  described above. 
Specifically, in the case of  TCS, the data were collected from objects WDS 11182+3132AB,   WDS 13491+2659, and WDS 08508+3504AB. 
Meanwhile, for  NOT, WDS J4514+1906AB and WDS J16289+1825AB were observed as well as the Trapezium region in the 2024 campaign. 
In all cases, pixel scales showed variations lower than 1 mas/pixel between the different objects and campaigns. 
All images shown in the paper are corrected to show north as up and east to the left. The axes of the images have a variation of less than one degree with respect to the axes on the sky in all measurements.

Finally, to apply the LIST algorithm, we performed the raw data reduction in two steps for each object individually. 
First, we selected the best  images of each set of the series corresponding to 1 per cent of the total, sufficient to provide reduced statistical errors but without including low quality images, as  demonstrated in \cite{Cagigal2016} and \cite{Cagigal2022}. The selection criterion was the brightest pixel, taking into consideration that the best PSF is the one with the brightest pixel and the worst, the one with the least bright pixel. 
Pixels with cosmic rays were discarded. Second, the images were cropped to $512\times 512$  pixels in those cases where they were of a larger size and the brightest pixel of each image was shifted to position (256, 256). 
After these two steps, we had a cube with 300-500 lucky images centred on the brightest pixel to which the LIST algorithm is applied via an \footnote{http://imagej.nih.gov/ijh}{ImageJ}  plug-in, requiring a post-processing time of a few seconds with a standard personal computer.

\section{LIST algorithm}
\label{sec:Peak}

Before introducing the LIST algorithm, we briefly describe the COELI and LISSA algorithms.
The COELI algorithm is based on the analysis of the temporal evolution of 
pixel intensities in a series of LI exposures  \citep{Cagigal2016}. The intensity of those pixels where a faint companion is located will fluctuate in phase with that of the main star intensity throughout the image series. 
However, the pixels containing incoherent speckles will fluctuate in the counterphase. The COELI technique relies on estimating the covariance between the intensity of the main star pixel and that of the remaining pixels throughout the LI cube to generate a 2D covariance map. This map highlights pixels where a companion is present, as they exhibit the highest covariance values.
The intensity of each pixel is given by :
\begin{equation}
    \text{C(x,y)} = \frac{<\text{i}_s.\text{i}(x,y)>-<\text{i}_s>.<\text{i}(x,y)>}{\sigma_s.\sigma(x,y)},
    \label{eq.eq1}
\end{equation}
where $i_s$  represents the intensity of the main star, $i(x,y)$ represents the pixel intensity, and $\sigma_s$ and $\sigma(x,y)$ are standard deviations \citep{Cagigal2017}. 

The LISSA algorithm is based on the statistics of the halo surrounding a host star in a series of short exposure images. The light intensity of the speckled  halo can be described by a modified Rician distribution \citep{Cagigal2022}. In the LI  technique, a coherent peak approximately twice the height of the surrounding speckles is needed to carry out an effective frame selection. In these conditions, the field in the 
halo area will consist of the sum of a constant phasor plus a series of random phasors with phase uniformly distributed in the interval $(-\pi,\pi)$. A similar situation, where a constant phasor is added to a series of 
random ones, can also be found in partial AO. In this case, it has been 
experimentally confirmed that a Rician distribution describes the speckle behaviour as 
well \citep{Canales1999}. In this context, we can define the parameter $r$ as:

\begin{equation}
    \text{r} = \frac{\text{I}_0}{\text{I}_n},
        \label{eq:eq2}
\end{equation}
where $I_n$  represents the average intensity of the random phasor sum alone and $I_0$ 
is the intensity of the constant phasor alone.

The expression for the halo intensity first moment is:

\begin{equation}
    \bar{I}= (1+r) \bar{I}_n.
        \label{eq:eq3}
\end{equation}

The STD and the signal-to-noise ratio are given by:

\begin{equation}
    \text{STD} = \bar{I}_n \sqrt{1+2r}
        \label{eq:eq4}
\end{equation}

and

\begin{equation}
    \frac{\bar{I}}{\text{STD}}= \frac{1+r}{\sqrt{1+2r}}.
        \label{eq:eq5}
\end{equation}

Equation \ref{eq:eq5} establishes a clear relationship between the halo intensity and its STD. Although the ratio of coherent to incoherent energy changes across the image plane, we use a single $r$ value. 
The $r$ value will also change with the actual seeing of the experiment. 
An average value of $r = 2$ generally produces an effective removing of the speckled halo. Nevertheless, this coefficient may be slightly different depending on the atmospheric conditions and the telescope.
Therefore, to estimate the halo intensity, we calculated the local STD in every frame and subtracted it, after multiplying by a factor (as described in Equation \ref{eq:eq4}). Thus, we were able to produce a new image where the halo has been removed, while maintaining the intensity of the companions

As previously mentioned, these algorithms were developed to detect faint companions in the area covered by the halo. The aim of this paper is to apply these techniques to short-exposure images in super-resolution mode 
which will allow us to resolve points even below the diffraction limit of the telescope.
The LIST algorithm has been optimised for detecting binary companions whose separation from their host stars is smaller than the distance to the first Airy ring
and we may consider it as a COLISSA optimization to highlight peaks around the centre of the image. Therefore LIST is composed of the following steps:

a)  The LI technique is applied to a set of short-exposure images, selecting the
 best frames based on their maximum peak intensity. We have experimentally checked that the best signal-to-noise ratio (S/N) was achieved for a number of frames ranging 300 to 500. This figure agrees well with that suggested by \cite{Fried1978} for a cube of 50000 frames. Therefore, a LI cube with the 500 best images will be generated. 

b)  Each image of the LI cube is to be centred in such a way that the pixel of maximum intensity
 will be at the centre of the image.

c)      The LISSA algorithm is applied in each image to reduce the speckled halo 
in the LI cube.

d)      A LF is applied in each image to highlight peaks around the centre of the image. 

e)      COELI is applied to estimate the normalized covariance (Pearson correlation) 
between the most intense peak of each image and the remaining pixels of the frame along the frame series. 

It is important to note that the resulting image cannot be used for measuring relative 
fluxes because the output of the algorithm is a covariance map. In other words, the 
images no longer represent real intensity images.

The principal difference between COLISSA and LIST is an intermediate Laplacian filter. Nevertheless, this relatively small addition allows LIST to achieve impressive results below the diffraction limit of 2-m class telescopes -- a regime where COLISSA is not effective.

\section{Validation of the technique}
\label{results}

\begin{figure*}[t]
\begin{center}
\centering
\includegraphics[ width=0.99\textwidth,angle=0]{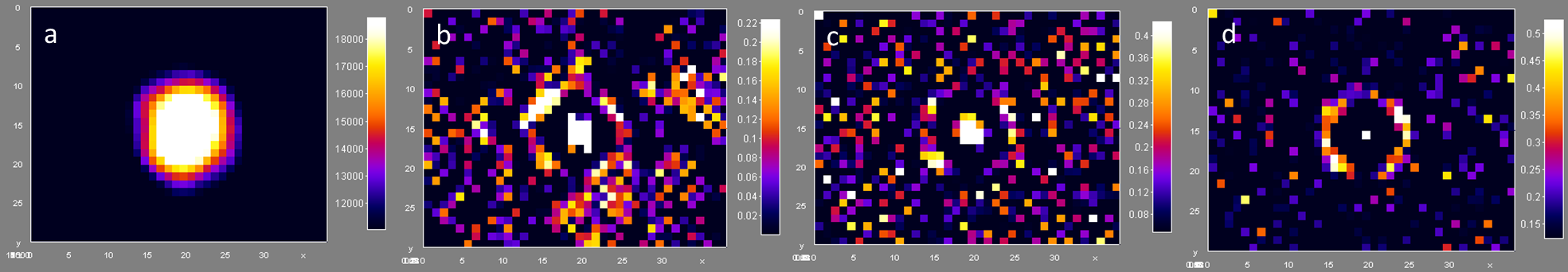}
\caption{Result of processing a stack of 500 images of the single star HIP 52457 using (a) the 
SAA algorithm, (b) the COLISSA algorithm, (c) the LF+COELI algorithm, and (d) the LIST algorithm. The field of view is 1.050 x 0.805 arcsec in all four figures}
\label{fig:fig2}
\end{center}
\end{figure*}

The performance of the algorithm was checked by applying it to the experimental data 
described in the previous section. In Sect. \ref{sec:simulations},  we describe how the algorithm was applied to simulations of visual binaries produced from single-star data. In Sect. \ref{sec:data}, observations of visual binaries with known
orbital elements are used and the expected ephemerides are contrasted with the results obtained after the application of the algorithm.

\subsection{Simulations of systems with multiplicity}
\label{sec:simulations}

 For the simulations, a previously selected cube of 500 frames was used to
apply LI to the data for the single star HIP 52457 observed with FC at 
TCS. This selected cube guarantees that the height of coherent peak is
approximately twice that of speckle mean intensity.

\begin{figure}
\begin{center}
        \includegraphics[width=\columnwidth]{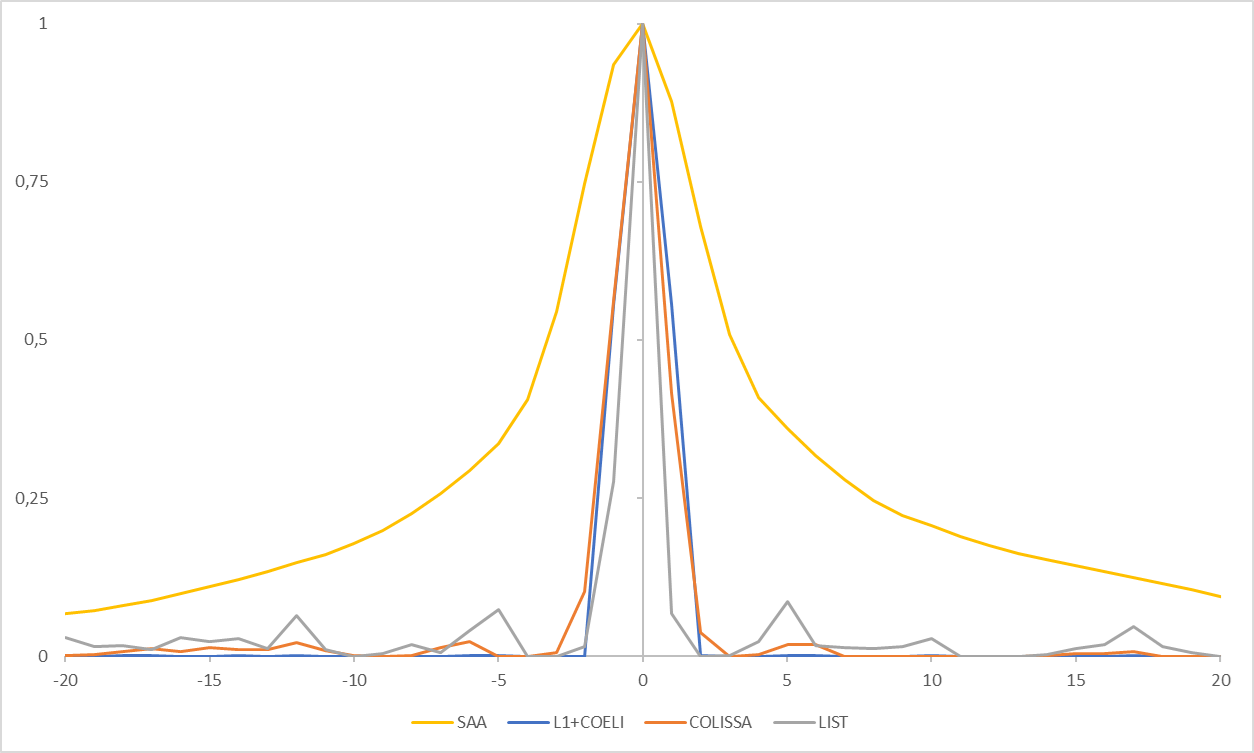}
    \caption{Cross-section of the result of processing a stack of 500 images of the single star HIP 52457 using the SAA algorithm (in yellow), LF+COELI (blue), COLISSA (orange), and the LIST algorithm (grey).}
    \label{fig:fig3}
\end{center}
\end{figure}

Figure~\ref{fig:fig2} shows the result of applying several techniques, (a) the Shift and Add algorithm (SAA), (b) the COLISSA algorithm, (c) the LF+COELI algorithm, and (d) the LIST algorithm.  Figure~\ref{fig:fig3} displays a cross-section for the four post-processing methods analysed in Figure~\ref{fig:fig2}.
It should be noted that for a seeing-limited image with the same instrumental set-up as for the previous analysis, we would obtain a Gaussian profile for the star with a full width half maximum (FWHM) of more than 25 pixels, depending on the night-time conditions. In the case of Figure~\ref{fig:fig2}, (a) where just the LI technique was
applied, we obtained a profile (yellow line in Figure~\ref{fig:fig3}) with a diffraction-limited central peak; in the this case, it is 4 pixels in the $I$-band at  TCS. Although this is an improvement over the case of an image limited by seeing, a residual halo corresponding to atmospheric turbulence is present, which cannot be corrected. In the profile corresponding to Figure~\ref{fig:fig2} (d), where the LIST algorithm was applied (grey line in Figure~\ref{fig:fig3}), the surrounding halo is suppressed and the central peak has a width of almost one pixel, providing a very narrow profile. The first Airy ring also appears, and the background between it and the central peak is practically null, thus increasing the probability of detection inside this area. In the profiles corresponding to Figures~\ref{fig:fig2} (b) and (c), where LF+COELI and COLISSA were applied (blue and orange lines in Figure~\ref{fig:fig3}, respectively), the central peak provides a wider profile, and the background is relevant. 

\begin{figure*}[t]
\begin{center}
\centering
\includegraphics[width=0.99\textwidth,angle=0]{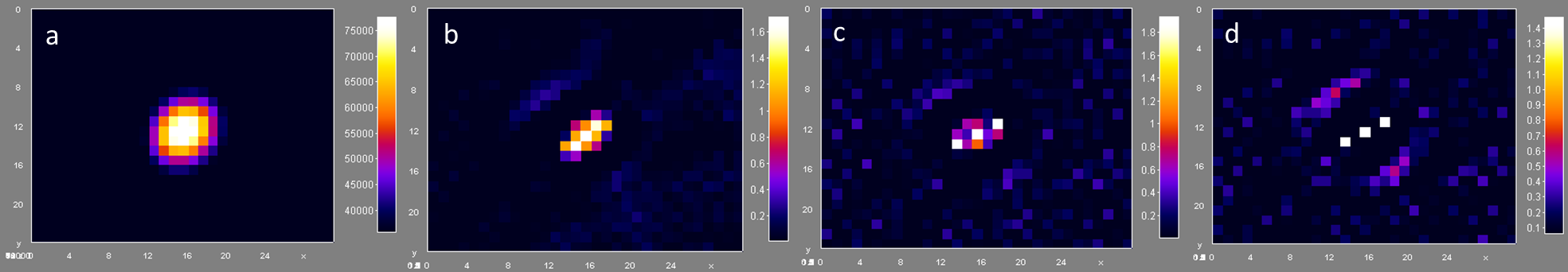}
\caption{Result of processing a stack of 500 images of a simulated binary object composed of a single star (HIP 52457) and a fake companion displaced 2.25 pixels using (a) the SAA algorithm, (b) the COLISSA algorithm, (c) the LF+COELI algorithm, and (d) the LIST algorithm. The field of view is 1.050\arcsec{}$\times$0.805\arcsec{} in all four figures.}
\label{fig:fig4}
\end{center}
\end{figure*}

Using the data for the single star HIP 52457 described previously, we simulated visual binaries at multiple separations, but always within the first ring of the Airy pattern. 
For the 1.52 m TCS, and taking into account the plate scale of the instrument in this configuration (35 mas/pixel), the first ring in the $I$-band appears at a distance of 5 pixels.
To generate simulated binary objects from the data for a single observed star, the cube 
previously selected with LI is duplicated and translated to a specific distance. The resulting cube was added to the original one 
resulting in a cube equivalent to that produced by two 
astronomical objects; for example, a binary star of similar intensity separated by the distance introduced. Before applying the algorithm, we re-centred all the frames in the cube and placed the pixel with the highest intensity at the centre of each frame. The brightest pixel in a frame may correspond to the main star or its companion, depending on the atmospheric perturbation. 
In Figure~\ref{fig:fig4}, we applied four different algorithms to synthetic data generated with a displacement of 2.25 pixels (2 pixels in X-direction and 1 pixel in Y-direction).
In  images (a) and (b), we applied the SAA and COLISSA algorithms, respectively. An elongation is seen in the direction, in which the displacement of the second object was applied, but the two components cannot be resolved. However, in the images (c) and (d) three well differentiated peaks appear, perfectly resolving the binary system when the LF+COELI algorithm or the LIST algorithm are applied.
Furthermore, Figure~\ref{fig:fig4} demonstrates that the LIST algorithm effectively clears the region within the first Airy ring, isolating the pixels corresponding to the host star and its binary companion.

\begin{figure}
\begin{center}
        \includegraphics[width=\columnwidth]{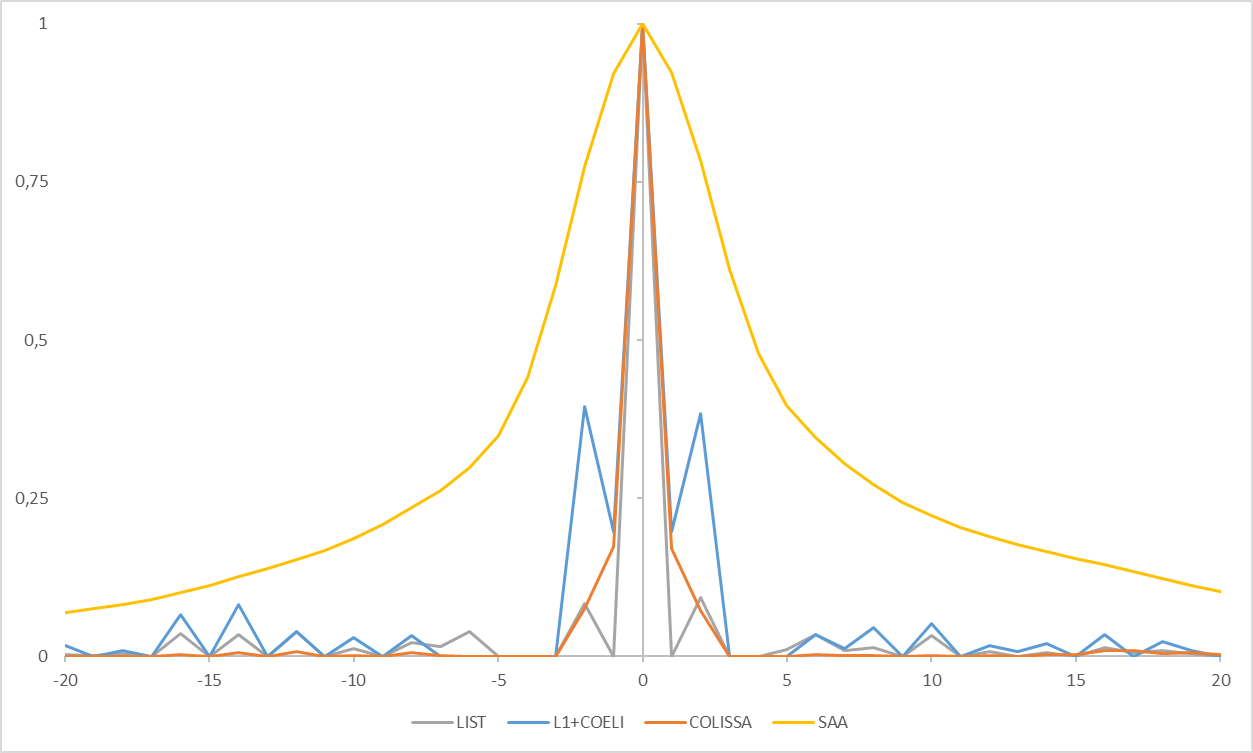}
    \caption{Cross-section of the result of processing a stack of 500 images of a simulated
 binary object composed of a single star (HIP 52457) and a fake companion displaced 2.25 pixels using
 the SAA algorithm (yellow, the LF+COELI algorithm (blue), the COLISSA algorithm (orange), and the LIST algorithm (grey).}
    \label{fig:fig5}
\end{center}
\end{figure}

Figure~\ref{fig:fig5} shows a cross-section in the direction of the displacement of Figure~\ref{fig:fig4}. 
The profile corresponding to the image after applying SAA (yellow) is similar to that for the single star, a diffraction-limited peak, and a halo corresponding to atmospheric turbulence. The profile after applying LIST (grey) shows a central peak and two secondary peaks corresponding to the simulated companion. The presence of three peaks is due to the movement of the maximum pixel in the re-centring process. In the case of an astronomical object of similar magnitude, the highest peak intensity may correspond to one object or another, depending on which one is randomly favoured by atmospheric turbulence.
In contrast, if the intensity of the companion is low enough, the re-centring process will provide a cube with all frames centred on the highest pixel of the main object, such that the centroiding will result in only two (rather than three) peaks after the algorithms are applied. 
The distance between the central and secondary peak corresponds to the separation between the main star and its companion (in this case: 2.25 pixels). Taking into account the characteristics of the instrument and the telescope, this separation is equivalent to 0.079\arcsec{}. Since the telescope's diffraction limit is 0.136\arcsec{} in the $I$-band, we have demonstrated that detection below the diffraction limit is feasible. 

\begin{figure*}[t]
\begin{center}
\centering
\includegraphics[ width=0.99\textwidth,angle=0]{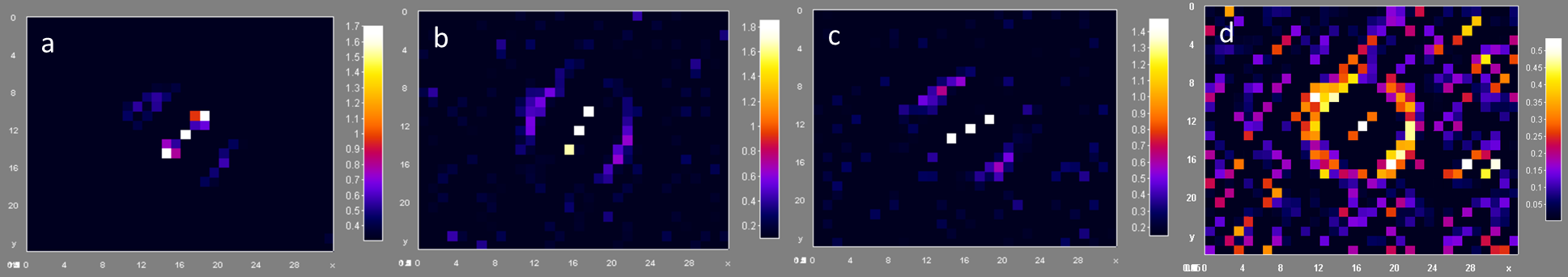}
\caption{Result of processing a stack of 500 images of a simulated binary object composed of the single star (HIP 52457) and a fake companion using the LIST algorithm. The fake companion is displaced by 
(a) 2.83 pixels, (b) 2.25 pixels, (c) 2.25 pixels, and (d) 1.41 pixels. These displacements correspond to the four cases detailed 
in Table \ref{tab:tab3}. The field of view is 1.050\arcsec{}$\times$0.805\arcsec{} in all four figures.}
\label{fig:fig6}
\end{center}
\end{figure*}

\begin{figure*} [t]
\begin{center}
\centering
\includegraphics[width=0.99\textwidth,angle=0]{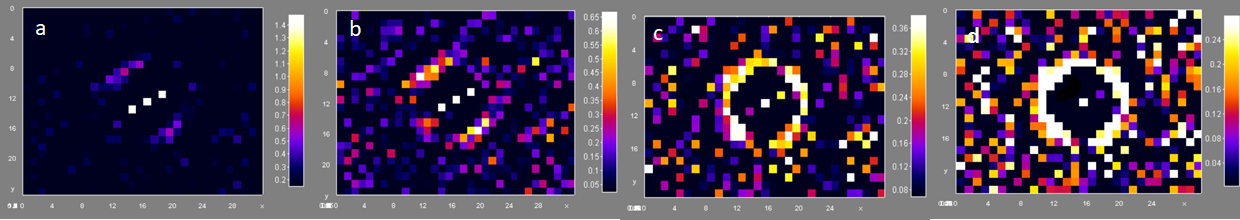}
\caption{Result of processing a stack of 500 images of a simulated binary object composed of a single star (HIP 52457) and a fake companion displaced 2.25 pixels using the LIST algorithm. The intensity  of the fake companion is divided by (a) 1, (b) 2, (c) 10 and (d) 25 with respect to the intensity of the 
main star. The field of view is 1.050\arcsec{}$\times$0.805\arcsec{} in all four figures.}
\label{fig:fig7}
\end{center}
\end{figure*}

After we checked that the detection of multiplicity of astronomical objects below the diffraction limit is possible, the distance between both stars was reduced in order to analyse the performance of the algorithm. 
Table~\ref{tab:tab3} lists the cases that have been analysed as a function of the separation between the two objects. In Figure~\ref{fig:fig6}, the cases corresponding to separations of 2.83 pixels, 2.25 pixels, and 1.41 pixels shown for the LIST algorithm, and in all cases detection is clearly possible. 

\begin{table}
        \centering
        \caption{Separations between fake 
companion and the host star for the different cases simulated. All units are in pixels.}
        \label{tab:tab3}
    \renewcommand{\arraystretch}{1.2} 
        \begin{tabular}{lccr} 
                \hline\hline
                Cases & X displacement & Y displacement & $\rho$\\
                \hline
                Case 1 & 2 & 2 & 2.83\\
        Case 2 & 1 & 2 & 2.25\\
                Case 3 & 2 & 1 & 2.25\\
        Case 4 & 1 & 1 & 1.41\\
                \hline
        \end{tabular}
\end{table}

We go on to analyse the algorithm's performance as a function of the difference in intensities between the two components of the simulated binary. Similarly to the aforementioned simulations, we generated  and recentred three cubes, where the intensity of a binary companion was reduced by a factor of 2, 10, and 25, respectively. Figure~\ref{fig:fig7} shows that LIST has  been applied to each cube and we observe that a detection is possible in all cases. 
Further simulations indicate that depending on the detection conditions, it could be achieved at intensity ratios as low as 50 (i.e.\ the binary companion is detectable when its intensity is 50 times lower than that of the host star). Below this threshold, detection becomes challenging because the energy of the first Airy ring surpasses that of the companion. 

\subsection{Application to visual binary stars}
\label{sec:data}

LI observations taken at the two telescopes have been analysed in cases where a visual orbit of the system predicted separations below the first Airy ring.
Figures \ref{fig:fig8} to  \ref{fig:fig11} show the images for each observation after applying LIST algorithm on  TCS (a) and on  NOT (b). In all images, north is up and east is to the left. 
For the four objects observed with both telescopes, when the post-processing based on the algorithm presented here is applied, in addition to the central pixel, other highlighted peaks appear in the inner area of the first Airy ring. These pixels are considered as possible detections of a system component. We chose this criterion because the LIST completely cleans the area inside the first ring with zero covariance value when analysing single stars. This results in any measurement of the contrast in this area being meaningless. Thus, even single pixels illuminated can be considered as system components.
A three-peaked profile appears for some of the objects because in those cases the components are of similar magnitudes ($\Delta m < 1$ mag). Depending on which component appears brightest in the individual images, a different component will be recentred. In the three-peaked profile, we only considered the components with higher covariance value.

\begin{table*}
        \centering
    \caption{Results obtained from LI observations after applying the LIST algorithm for astrometric measurements of the separation and position angle between the detected components. }

\label{tab:astrometric_result}

\setlength{\tabcolsep}{15pt} 
\renewcommand{\arraystretch}{1.2} 
\begin{tabular}{cccccccc}
 
  \hline\hline
                WDS                         &  Telescope  &    $\overline{\rho}$    & Jaccard & $\overline{\theta}$  & Jaccard \\
        ($\alpha$,$\delta$,J2000.0) &             &      (")     &   index      & ($\degr$) & index  \\
        (1)                         &    (2)      &      (3)     & (4)     &  (5)      &   (6)       \\
\hline
08585+3548 &  NOT  &  0.098  &  0.08 & 42   & 0.29   \\
08585+3548 &  TCS  &  0.099  &  0.08 & 45   & 0.29   \\
10083+3136 &  NOT  &  0.092  &  0.43 & 168  & 0.13   \\
10083+3136 &  TCS  &  0.085  &  0.25 &  343 & 0.10   \\
10116+1321 &  NOT  &  0.092  &  0.64 &  280 & 0.21  \\ 
10116+1321 &  TCS  &  0.085  &  0.43 &  270 & 0.36  \\
15245+3723 &  NOT  &  0.101  &  0.07 &  268 & 0.14  \\
15245+3723 &  TCS  &  0.080  &  0 &  279 & 0.09  \\

\hline
\end{tabular}
\tablefoot{Columns: (1) is the WDS number and (2) the telescope used. (3) shows the angular separation measured in arcseconds and (4) the Jaccard index for this variable.  (5) the position angle between the two components obtained in the images, and (6) the Jaccard index for this variable.
}
\end{table*}

Based on the images after the application of the LIST analysis, in Table \ref{tab:astrometric_result} we present
 the results obtained with the astrometry of the studied systems.
The columns present the following data: (1)   WDS number (this also gives the right ascension and declination of the object in J2000.0 coordinates); (2)  telescope used in the observations; (3)   observed separation $\overline{\rho}$ in arcseconds; (5)  observed position angle, $\overline{\theta}$, of the secondary star relative to the primary, with north through east defining the positive sense; and (4) and (6) Jaccard index for separation and position angle for each object, defined below. It should be noted that in the case of WDS 10083+3136 observed by  NOT, the angle obtained is in the complementary quadrant with respect to the expected one. This is because the magnitude difference between the two components is small ($< 0.5$ mag) and the brighter pixel can oscillate between the two depending on the atmosphere. In this case, 180 degrees was added to allow comparison with the expected ephemerides. 

The position of the main object is determined by the central pixel, due to the fact that the images were re-centred to the maximum pixel during the analysis. 
To calculate the position of the detected companion, we considered 
the group of pixels with signal around the pixel with the highest covariance value.

Using the $n$ highlighted pixels for the second component in the region of detection with coordinates $(x_{i},y_{i})$, the position of the companion $(x_{2},y_{2})$ is computed as:

\begin{equation}
     (x_{2},y_{2}) = (\sum_{i=1}^{n} w_{i} \cdot x_{i},\sum_{i=1}^{n} w_{i} \cdot y_{i}),
        \label{eq:eq6}
\end{equation}
where $w_{i}$ is the normalised weight of each pixel with signal 
$s_{i}$ defined as $w_{i}= s_{i} / \sum_{i=1}^{n} s_{i}$.

Having computed the positions of the host star and the companion, the apparent angular separation in arcseconds is given by:

\begin{equation}
     \rho =  p \cdot \sqrt{(x_{2}-x_{1})^2 + (y_{2}-y_{1})^2 },
        \label{eq:eq7}
\end{equation}
where $p$ is the plate scale of the detector in each configuration.
The  position angle ($\theta$) in radians, counted counterclockwise from north up and east left, is calculated to be:

\begin{equation}
     \theta = \arctan \left(\frac{x_{1}-x_{2}}{y_{2}-y_{1}} \right).
        \label{eq:eq8}
\end{equation}

The astrometric accuracy  associated with our measurements is obtained by taking two things into consideration.
On the one hand, the determination of the coordinates of the highlighted pixels has  an associated error of half a pixel ($x_{i}\pm0.5$,$y_{i}\pm0.5$). 
We have been conservative with this value, as we have to bear in mind that we are not working with intensity images, but with images that represent the covariance map of our analysis.
On the other hand, the weighted standard deviation of the coordinates $(x_{2},y_{2})$ is used when  more than one highlighted pixel is available for use in the calculation. 
Thus, the average accuracy obtained in the measurement of the angular separation for the four systems observed with  NOT is  $16 \pm 2$ mas and with  TCS, it is  $20 \pm 1$ mas.
In both cases, this error corresponds to roughly two thirds of the plate scale in each configuration. 
Regarding the average accuracy of the position angle, measurement is  
 $9.5 \pm 0.3 ^o$ for  NOT and  $11 \pm 2 ^o$ for  TCS.
These values make sense because of the limited sampling we have within the area inside the first Airy ring.

\begin{figure}[H]
\includegraphics[width=\columnwidth]{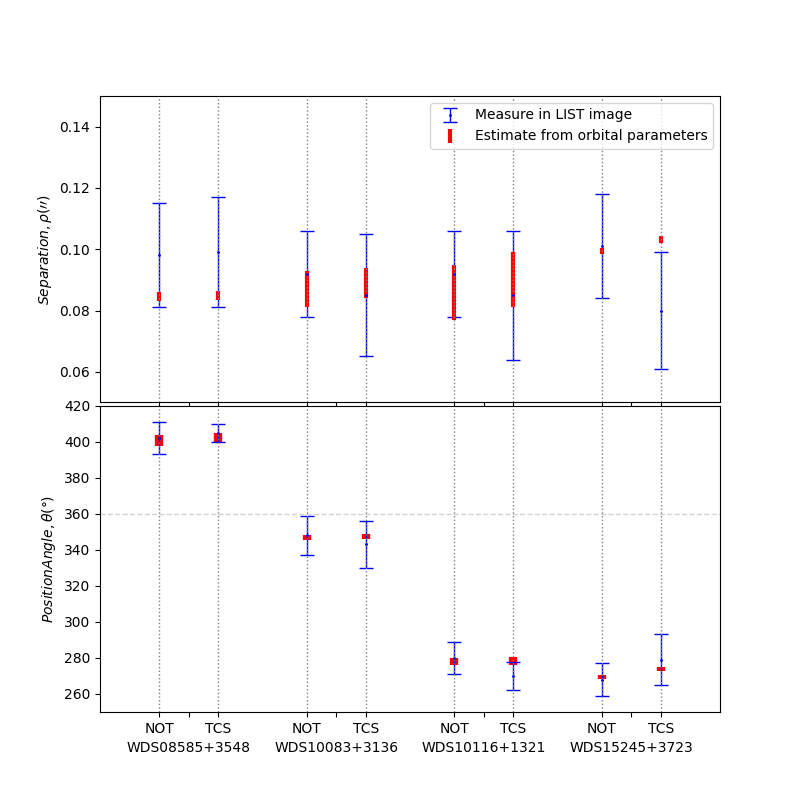}
\caption{Comparison of the ephemeris values measured in the LIST images (blue) with those calculated from the known orbital parameters (red). The top plot represents the projected separation variable, while the bottom plot represents the positional angle between the two components. For clarity, 360 degrees have been added for the first object. On the x-axis, each of the binaries and the telescopes used for the observations are specified.}
\label{fig:fig12}
\end{figure}

\begin{figure} [h]
    \includegraphics[width=0.92\columnwidth]{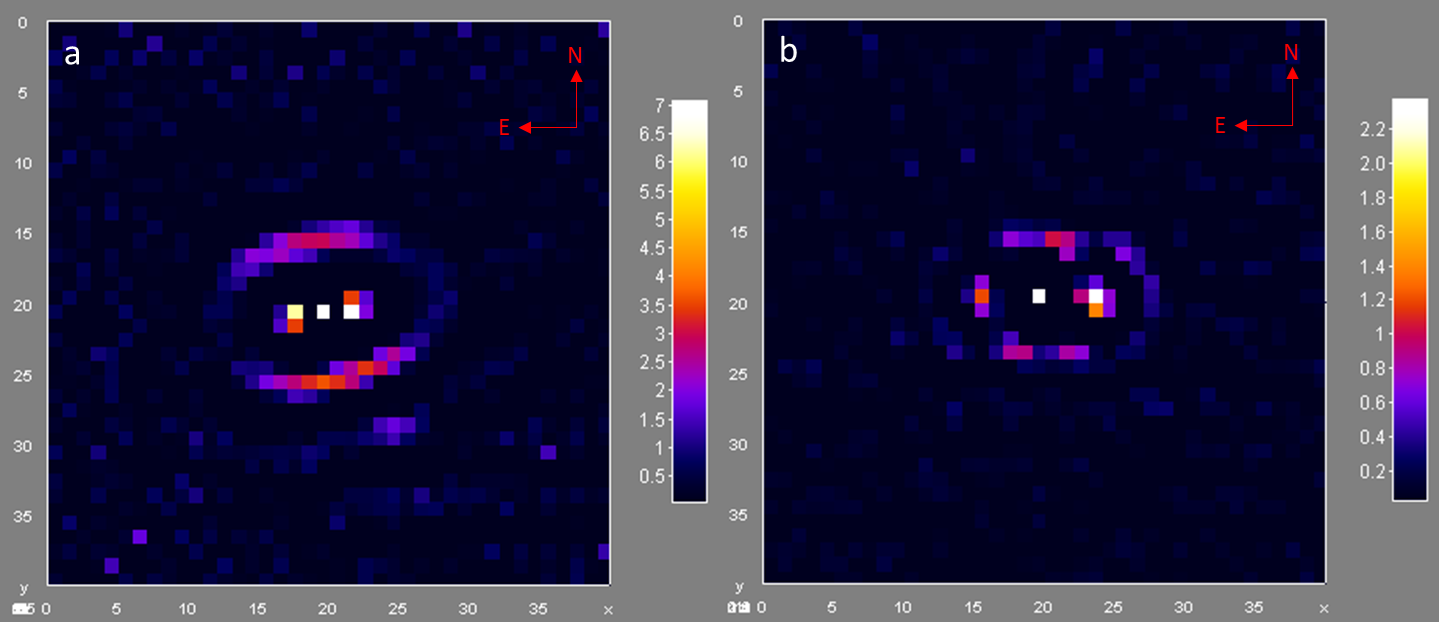}
    \caption{Result of processing a stack of 500 images of the object WDS15245+3723 using LIST algorithm on (a) TCS and (b) NOT. The fields of view are: 1.4\arcsec{}$\times$1.4\arcsec{} and 1.0\arcsec{}$\times$1.0\arcsec{} for  TCS and NOT, respectively}
    \label{fig:fig8}
\end{figure}

\begin{figure} [h]
        \includegraphics[width=0.92\columnwidth]{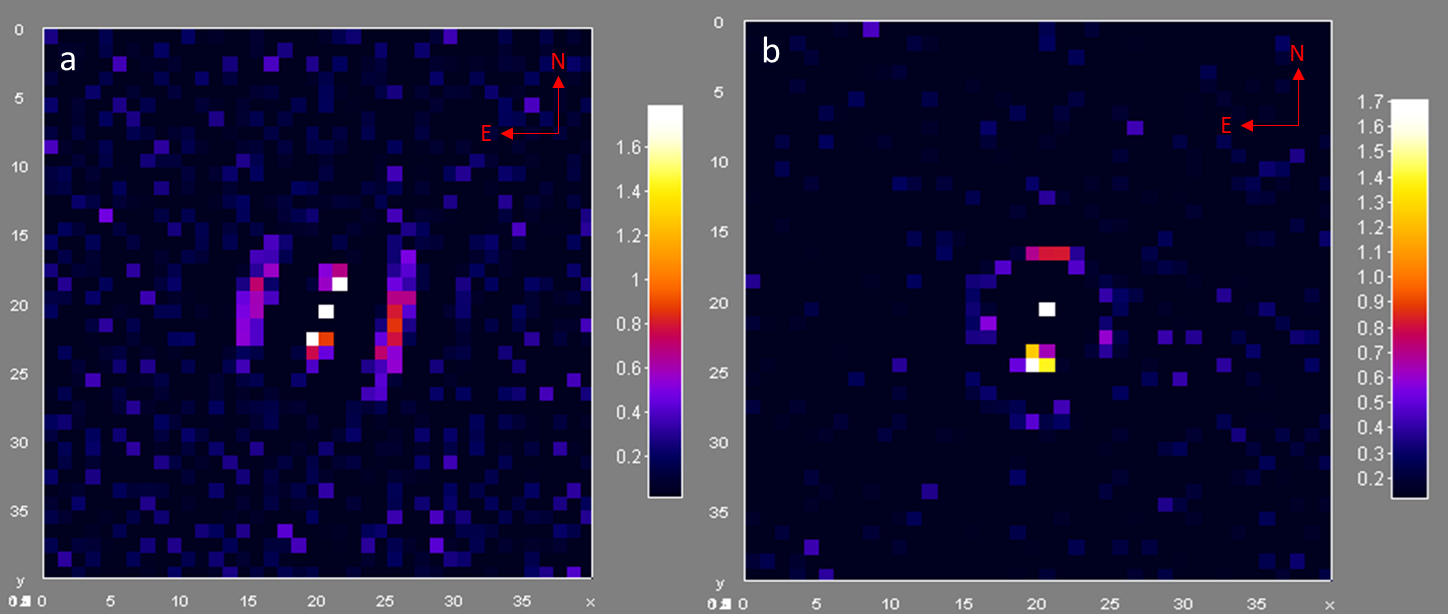}
    \caption{Result of processing a stack of 500 images of the object WDS10083+3186 using LIST algorithm on (a) TCS and (b) NOT. The field of view are 1.4\arcsec{}$\times$1.4\arcsec{} and 1.0\arcsec{}$\times$1.0\arcsec{} for  TCS and NOTm respectively}
    \label{fig:fig9}
\end{figure}

\begin{figure} [h]
        \includegraphics[width=0.92\columnwidth]{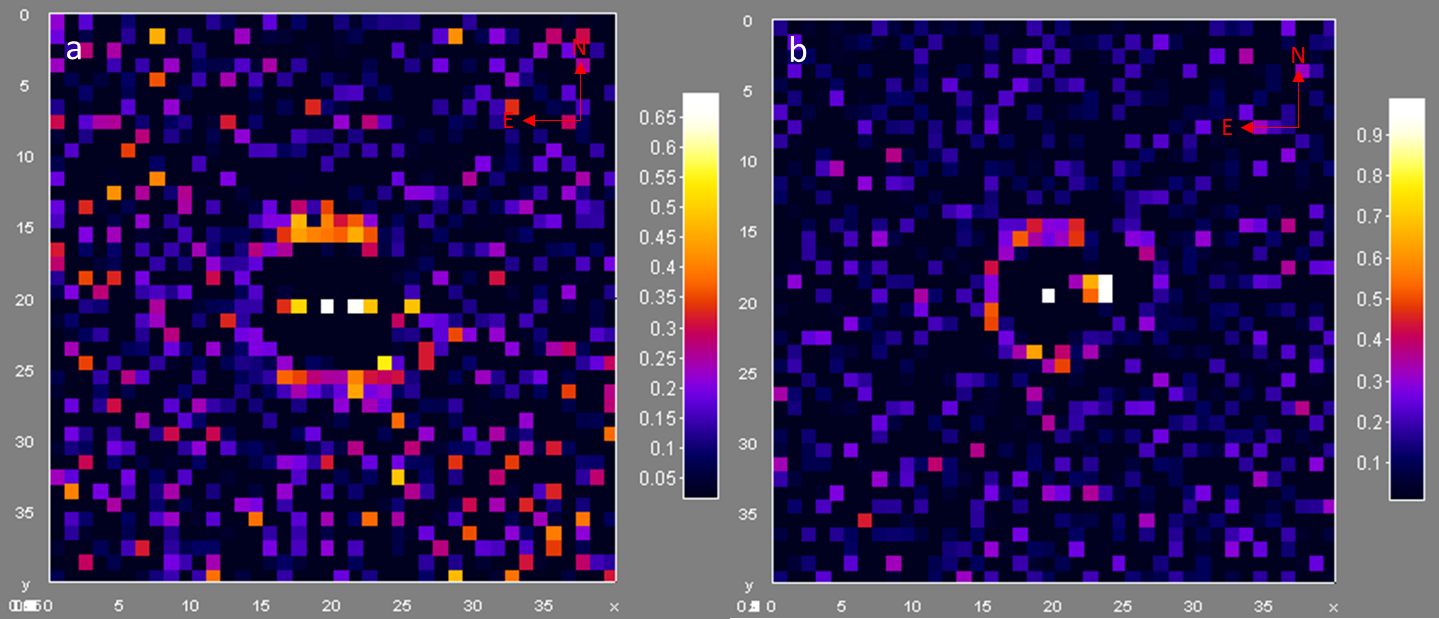}
    \caption{Result of processing a stack of 500 images of the object WDS10116+1321 using LIST algorithm on (a) TCS and (b) NOT. The fields of view are: 1.4\arcsec{}$\times$1.4\arcsec{} and 1.0\arcsec{}$\times$1.0\arcsec{} for  TCS and NOTm respectively}
    \label{fig:fig10}
\end{figure}

\begin{figure} [H]
        \includegraphics[width=0.92\columnwidth]{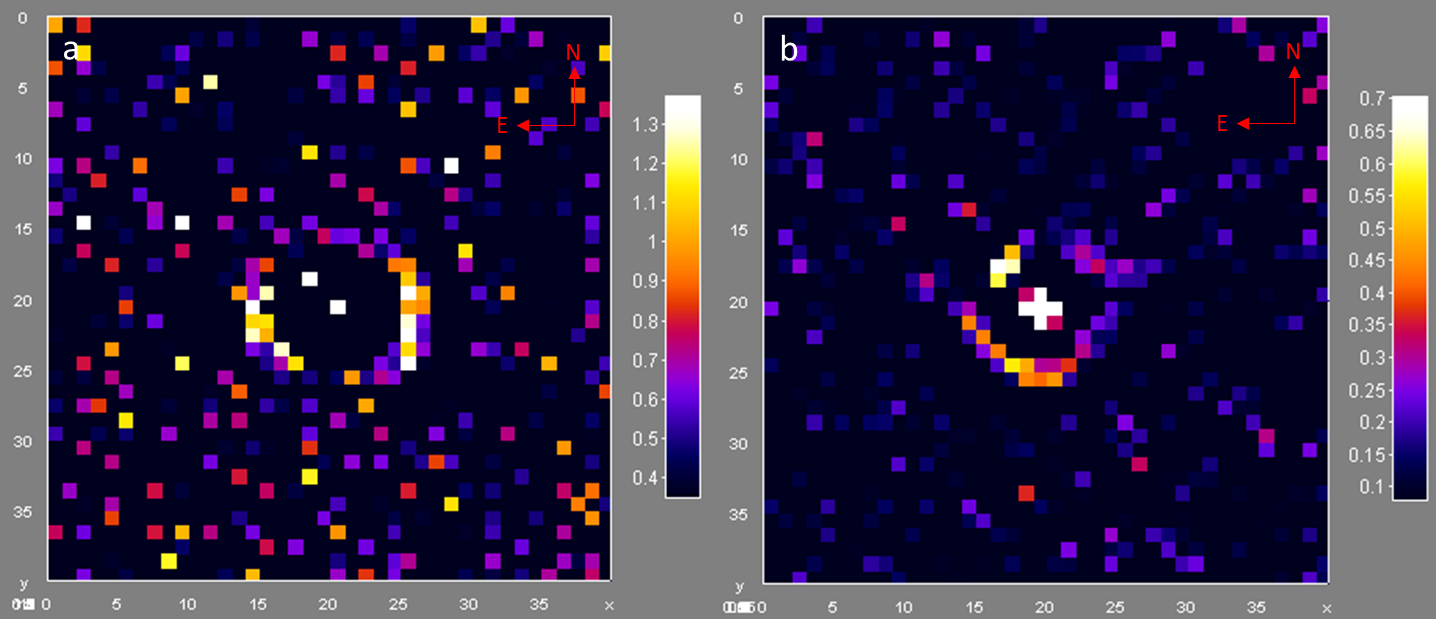}
    \caption{Result of processing a stack of 500 images of the object WDS08585+3548 using LIST algorithm on (a) TCS and (b) NOT. The fields of view are: 1.4\arcsec{}$\times$1.4\arcsec{} and 1.0\arcsec{}$\times$1.0\arcsec{} for  TCS and NOT, respectively}
    \label{fig:fig11}
\end{figure}

Using the uncertainties of the orbital parameters provided in the Sixth Catalogue, we calculated the ranges of separation and position angle where these variables can be found at the time of observation. Figure \ref{fig:fig12} shows the comparison of these ranges with the ranges obtained from our measurements and the calculated uncertainties discussed above for each object. 
The x-axis shows the four binaries observed and the corresponding NOT and TCS telescope observations for each object are marked with vertical dotted lines.
In the graphic below, the y-axis represents the angle of position between the components. To ensure that the scale of the y-axis would allow for a more detailed appreciation of the data presented, 360 degrees were added to the angle of the first object (marked by a horizontal dashed line).
In the graphic above, the y-axis represents the separation between the two components in arcseconds.
In both plots, the data in blue correspond to those measured in the LIST images, while the red ones are the expected ephemerides calculated from the known orbital elements.
For both variables an intersection is found, to a greater or lesser amount, to lie between the ranges calculated and measured in the images. WDS 15245+3723  (observed with  TCS) is the only exception where the value of the projected separation is still close, but do not overlap. This may be due to the limited spatial resolution of the camera (c.f.\ the expected angular separation).

The Jaccard index is used to evaluate, in a quantitative way, the coincidence of the calculated  variables. This index, also known as the Jaccard similarity coefficient, is a statistical measurement used to quantify the similarity between two sets. It is defined as the size of the intersection of the sets divided by the size of the union of the sets as:

\begin{equation}
    J(A,B)=\frac{A\cap B}{A \cup B},
        \label{eq:eq10}
\end{equation}
with values ranging from 0 (indicating no similarity) to 1 (indicating 
complete similarity). 

Table~\ref{tab:astrometric_result} also includes the Jaccard index for both the separation and position angle for each object, columns (4) and (6), respectively. For all cases, it is greater than 0, except in the case discussed above for WDS 15245+3723 (observed with TCS), where there is no overlap. It is also observed that the index is small (around 0.1) for some of the objects for both parameters. This is because the errors of the ephemerides calculated from the orbital elements are of the order of $1\%$; whereas  for the calculation based on the images, the errors are 15--20\% and 5--10\%  for the separation and position angle, respectively. In the other cases, where the errors are of the same order, the Jaccard index is larger than 0.25. 
Hence, in all but one of the comparisons, the measurements agree with the ephemerides to within the errors. This observational test of known binaries with FC shows that the LIST processing has resolved binaries with separations similar or below the diffraction limit of the two telescopes. This has allowed us to measure separations and position angles that generally agree with the ephemerides to within the estimated errors.

Although the LIST algorithm can detect companions in the super-resolution regime, it is not able to provide photometry. To overcome this drawback we developed an algorithm, named PHOTO-LIST, based on the use of LIST and LISSA algorithms. It consists of the following steps:

1. Select a 500 frame cube and apply LIST algorithm to obtain the covariance map image and thus be able to detect a possible candidate.

2. The resulting map is binarized, so that pixels below a threshold are set to 0 and the rest to 1.

3. Select the same 500 frame cube, apply the LISSA algorithm to remove the image speckled halo, and obtain an average LISSA image.

4. Multiply the average LISSA image by the binarized LIST image.

\begin{table} [H]
        \centering
        \caption{Measurements of the differential photometry, $\Delta$m,  of the four studied systems. }
        \label{tab:delta}
   \setlength{\tabcolsep}{4pt} 
    \renewcommand{\arraystretch}{1.3} 
        \begin{tabular}{c|cc|cc} 
                \hline\hline
                WDS & Reference & $\Delta m$ & $\Delta m_{NOT}$ & $\Delta m_{TCS}$\\
        ($\alpha$,$\delta$,J2000.0) &                &      &  \\
                 (1)             &    (2)         & (3)  & (4)  &  (5)  \\
     \hline
        08585+3548 & \citealt{2020AJ....159..233H} & 1.51 & 1.57 & 1.75\\
                10083+3136 & \citealt{2020AJ....159..233H} & 0.72 & 0.78 & 0.59\\
        10116+1321 & \citealt{2020AJ....159..233H}   & 1.19 & 1.10 & 0.71 \\
        15245+3723 & \citealt{2014ARep...58..835K} & 0.41 & 0.53 & 0.62\\

                \hline
        \end{tabular}
\tablefoot{The discoverer designation for each object,  photometric measurement used as a reference, and  measurements obtained at each telescope  are shown.  The specifications for each case are given in the main text. Note: the reference values have been obtained using filters with a slightly different centre value and bandwidth than the ones used in our experiments
}
\end{table} 
This results in an intensity image, rather than a covariance map (as with LIST).  In this way, we can  attempt to measure the differential photometry between the two components.  Table~\ref{tab:delta} presents the following columns: (1)  WDS number; (2)  details of the references where  magnitude difference, $\Delta$m, of each companion was taken from the  listed in column (3) and 
the measurements from \cite{2020AJ....159..233H} are at a wavelength of 880 nm, while for \cite{2014ARep...58..835K} it is 800 nm; and (4) and (5)  $\Delta$m  measurements obtained for the objects observed in  NOT and TCS, respectively. In our case, the wavelength $\lambda$ is 825 nm.  Also, the bandwidth in each configuration can vary depending on the width of the filters as well as the quantum efficiency of the detectors used.  Therefore, it should be noted that this comparison serves to validate the presented method and to estimate the accuracy of our measurements,  but there may be slight differences between them for the reasons mentioned above. In fact, it can be seen that in the case of  NOT, the residual of the comparison with the reference measurements is less than 0.1. In the case of  TCS there is a greater discrepancy, with the largest residual being 0.48. This may be due to the difference in the quality of the images taken with the two telescopes or also to the fact that the TCS measurements were made below the diffraction limit, which may be less sensitive than in areas above the first Airy ring.

As a final remark, it should be noted that in an observation of this nature where the telescope is pushed beyond its own limits, it is very important to rule out adverse phenomena of other nature that may distort the measurement. There are several possible causes of interference. Besides the main cause of distortion, which is the fluctuation of the refractive index of the atmosphere, there can also be the effect of chromatic dispersion, the color differences between the primary and the companion (and even the wind). On the other hand, problems can also arise from the telescope that limit the quality of the images, such as possible residual aberrations, telescope guiding, the detector being out of focus, and other smaller effects. Therefore, stable meteorological conditions and a meticulous data collection process as well as a subsequent analysis of the quality of the images before applying the algorithm are necessary to aid our understanding of the possible effects that may have contributed to the images.

\section{Discussions and conclusions}

In this study, we present an algorithm to analyse images of LI observations identifying  companions to astronomical objects, specifically focusing on binary stars. In particular, we employed a super-resolution imaging approach that operates below the diffraction limit of the telescope, as defined by the Rayleigh criterion.

For this purpose, we used the FC instrument mounted on two telescopes:\ 1.5m TCS and  2.5m NOT. In both configurations, the condition $D/r_{0} <$ 8 is fulfilled, which is necessary to have a good signal at the coherent peak with respect to the speckled halo. This condition does not have to be fulfilled with other techniques used in similar works, such as speckle imaging, where telescopes with larger diameters are usually employed. The use of telescopes of modest sizes has the advantage that they are not so over-subscribed, thereby offering greater access to observing time and the ability  to carry out routine observations and long-period programmes.

Referred to as LIST, the algorithm essentially combines two pre-existing techniques, COELI and LISSA introducing a LF between them. These techniques were developed to suppress the influence of the incoherent light halo generated in the temporal average of LI-selected images.
This halo spans an area ranging from the inner radius of $1.22\lambda/D$ to the outer radius of $1.22\lambda/r_0$.
The algorithm's optimization, as detailed in this article, is specifically tuned for the region below the first Airy ring ($1.63\lambda/D$), resulting in a reduction of the FWHM of the Airy disk to less than one pixel. 
In order to test the performance of the algorithm,  binary systems were chosen for the observation whose orbital parameters predict a separation between components in the super-resolution regime in TCS, being close to the diffraction limit in the case of  NOT. This has served to rule out systematic effects and errors in the application of our method and confirm the veracity of the detection of components in the aforementioned area.

Therefore, the LIST algorithm extends the range of separations that can be measured in binary systems using LI observations at visible wavelengths; in particular, at 825 nm. In the cases described, objects have been resolved at almost half the resolution limit of the system used. This range would increase the measurements of projected angular separation below 0.15\arcsec{} down to approximately 0.05\arcsec{} using only the simple LI technique and post-processing analysis.

This is in contrast to the greater technological resources and larger telescopes used with other techniques to cover similar ranges. 
However, the uncertainties in the separations and the position angle achieved in our experiment are larger than those obtained with speckle imaging techniques on sub-diffraction limited images.
It should be noted that both the separations that can be resolved and the error associated with them using our technique are limited by the plate scale defined in each configuration. However, it should be considered that the constant reduction of the plate scale leads to a lower S/N in the images, so a compromise between the two must be reached in order to obtain an optimum configuration.

This capabilities of LIST have been substantiated through two key approaches. First, by creating synthetic observations of visual binaries,  with different angular separations and intensities, based on the real data of a single star. With this test, we found that the highest S/N in the area defined by the first Airy ring is obtained with LIST when compared with the other algorithms reviewed.
Second, we validated our methodology with real objects -- binary stars with well known orbits and separations -- finding that the projected angular separations and position angles determined using LIST are in agreement within the estimated errors of the known ephemerides.

In this preliminary test of the LIST algorithm, we have observed binaries where the brighter component is characteried by $m_{V}<7$. However, in routine FC observations, with the current configuration on both telescopes, it has been estimated that  to be able to select the images  and re-centre them appropriately, the limiting magnitude of at least one object is around $m_{V}=11$ (although it specifically depends on the type of star). It would be possible to go deeper if the weather conditions were optimal. Regarding the relative photometry, which (in principle) was not performed by the presented algorithm, the PHOTO-LIST analysis based on the techniques summarised in the paper has been developed. Thus, an attempt to measure the magnitude difference, $\Delta$m, for the studied systems has been made. The system observed with the largest difference is estimated to be around 1.5 magnitudes, although simulations have recovered components with $\Delta$m around 4. However, the photometric measurement degrades as the companion becomes fainter and the $\Delta$m in these cases is less accurate. We find that the LIST algorithm is less sensitive in terms of both limiting magnitude and magnitude difference than analyses above the diffraction limit. Future observations will attempt to reach the limiting values of the presented configurations, using this technique.

To the best of our knowledge, this study marks the first successful resolution of objects at  sub-diffraction-limited projected angular distances in the optical range using medium-sized telescopes with the LI technique. 
For an idea of the range of our technique with the configurations we  employed, for a binary system at 100 pc, an object at 0.05" corresponds to a distance of 5 AU. This information has the potential to increase the efficiency of other studies, such as large surveys carried out with more complex experiments (e.g. at a long-baseline optical interferometer).
This achievement underscores the significance of our method as an innovative tool for analysing multiplicity in astronomical objects, such as binary stars with intermediate separations, while requiring fewer resources than  the techniques  typically employed in these scenarios.


\begin{acknowledgements}
This article is partly based on observations made at the Telescopio Carlos S\'anchez operated on the island of Tenerife by the Instituto de Astrof\'isica de Canarias in the Spanish Observatorio del Teide. 
Partly based on observations made with the Nordic Optical Telescope, 
owned in collaboration by the University of Turku and Aarhus University, 
and operated jointly by Aarhus University, the University of Turku and 
the University of Oslo, representing Denmark, Finland and Norway, the 
University of Iceland and Stockholm University at the Observatorio del 
Roque de los Muchachos, La Palma, Spain, of the Instituto de Astrof\'isica 
de Canarias.
The observations made for this article have been partially planned with the use of ``Stelle Doppie" (the site www.stelledoppie.it).

DJ acknowledges support from the Agencia Estatal de Investigaci\'on del Ministerio de Ciencia, Innovaci\'on y Universidades (MCIU/AEI) under grant ``Nebulosas planetarias como clave para comprender la evoluci\'on de estrellas binarias'' and the European Regional Development Fund (ERDF) with reference PID-2022-136653NA-I00 (DOI:10.13039/501100011033). DJ also acknowledges support from the Agencia Estatal de Investigaci\'on del Ministerio de Ciencia, Innovaci\'on y Universidades (MCIU/AEI) under grant ``Revolucionando el conocimiento de la evoluci\'on de estrellas poco masivas'' and the the European Union NextGenerationEU/PRTR with reference CNS2023-143910 (DOI:10.13039/501100011033).
\end{acknowledgements}

\bibliographystyle{aa}
\bibliography{LIST_AA.bib}

\end{document}